\documentclass[aps,twocolumn,prl,floatfix,showpacs,superscriptaddress]{revtex4}

\usepackage{amsmath}
\usepackage{amssymb}
\usepackage{amsfonts}
\usepackage[pdftex]{graphicx}
\usepackage{units}
\usepackage[usenames]{color}
\usepackage{dcolumn}
\usepackage{bm}
\usepackage{float}

\newcommand{\ket}[1]{\mid\!\! #1 \rangle}

\newcommand{\cone}{\mathrm{i}}

\newcommand{\ion}[1]{\ensuremath{\mathit{#1}}}

\begin{document}
\title{Pressure-induced Topological Phase Transitions in Rock-salt Chalcogenides}

\author{Paolo Barone}
\affiliation{Consiglio Nazionale delle Ricerche (CNR-SPIN), I-67100 L'Aquila, Italy}

\author{Tom\'a\v{s} Rauch}
\affiliation{Institut f\"ur Physik, Martin-Luther-Universit\"at Halle-Wittenberg, D-06099 Halle (Saale), Germany}

\author{Domenico Di Sante}
\affiliation{Consiglio Nazionale delle Ricerche (CNR-SPIN), I-67100 L'Aquila, Italy}
\affiliation{Department of Physical and Chemical Sciences, University of L'Aquila, Via Vetoio 10, I-67010 L'Aquila, Italy}

\author{J\"urgen Henk}
\affiliation{Institut f\"ur Physik, Martin-Luther-Universit\"at Halle-Wittenberg, D-06099 Halle (Saale), Germany}

\author{Ingrid Mertig}
\affiliation{Institut f\"ur Physik, Martin-Luther-Universit\"at Halle-Wittenberg, D-06099 Halle (Saale), Germany}

\author{Silvia Picozzi}
\affiliation{Consiglio Nazionale delle Ricerche (CNR-SPIN), I-67100 L'Aquila, Italy}


\begin{abstract}
By means of a comprehensive theoretical investigation, we show that external pressure can induce topological phase transitions in IV-VI semiconducting chalcogenides with rock-salt structure.
These materials satisfy mirror symmetries that are needed to sustain topologically protected surface states, at variance with time-reversal symmetry responsible for gapless edge states in $\mathcal{Z}_{2}$ topological insulators.
The band inversions at high-symmetry points in the Brillouin zone that are related by mirror symmetry,
are brought about by an ``asymmetric'' hybridization between cation and anion $sp$ orbitals. By working
out the microscopic conditions to be fulfilled in order to maximize this hybridization, we identify
materials in the rock-salt chalcogenide class that are prone to undergo a topological phase transition
induced by pressure and/or alloying. Our model analysis is fully comfirmed by complementary advanced \textit{first-principles} calculations and \textit{ab initio}-based tight-binding simulations.
\end{abstract}


\maketitle

Semiconducting chalcogenides \ion{CA}, in which the cation \ion{C} is an element of the group IV (Ge, Sn, Pb) and the anion \ion{A} is an element of the group VI (S, Se, Te),  represent an attractive class of materials, due to their unique structural and electronic properties. The most interesting compounds in this class --- SnTe, GeTe, and the lead chalcogenides Pb\ion{A} --- display rock-salt structure  \cite{Abrikosov68}. Both SnTe and GeTe have been long known for their ferroelectric properties in the low-temperature distorted structure \cite{Lines77}, whereas lead chalcogenides possess potential relevance for thermoelectric and optoelectronic applications \cite{Khokhlov03}. Their outstanding properties have been characterized by a variety of experimental techniques \cite{Khokhlov03,Hoelher83}. In parallel, a large number of theoretical investigations --- carried out with different methods ranging from \textit{ab initio}-based tight-binding to density-functional theory calculations --- have addressed their peculiar electronic structure \cite{Mitchell66,Lent86,Lach-hab02,Wei97,Hummer07,Svane10}. Interest in this material class has been renewed because of relativistic effects that may prove relevant for future spintronic devices. For example, a giant Rashba effect has been predicted in the ferroelectric phase of bulk GeTe \cite{DiSante13}. Furthermore,  SnTe in its undistorted structure has been predicted to be a topological crystalline insulator (TCI) \cite{Hsieh12}; its spin-polarized surface states have been confirmed by photoelectron spectroscopy later on \cite{Tanaka12}. These observations suggest to investigate the possibility of topological phase transitions and their microscopic conditions in this class of narrow band gap semiconductors.

\begin{figure}
 \centering
 \includegraphics[width=\columnwidth]{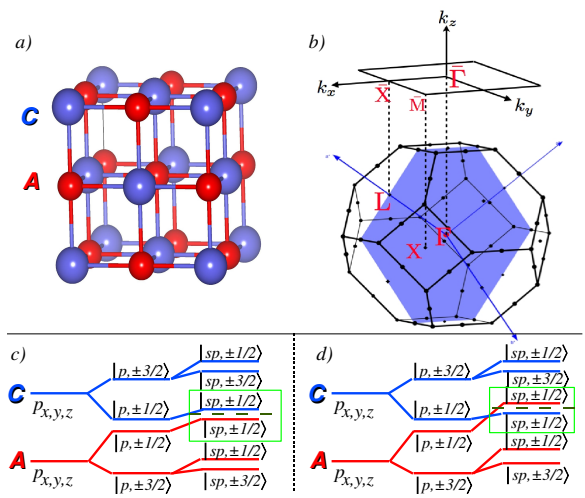}
 \caption{(Color online) Crystal structure of rock-salt chalcogenides \ion{CA} (a) and the fcc Brillouin zone (b), showing one mirror plane (green) containing the high-symmetry points $\Gamma$, L, and X as well as their projection onto the [001] surface. Schematic representation of the band structure at the L point in the topologically trivial phase (c) and upon pressure-induced enhancement of $sp$ hybridization (d). Dashed lines separate occupied from unoccupied energy levels.}
 \label{fig:crystal}
\end{figure}

In the $\mathcal{Z}_{2}$ class of topological insulators, time-reversal symmetry ensures topologically protected edge states\cite{Hasan10,Hasan11}. An odd number of band inversions distinguishes a $\mathcal{Z}_{2}$ topological insulator from a conventional band insulator: the former shows an odd number of Dirac cones pinned at time-reversal-invariant momenta (TRIM),  as found for example in Bi$_{1-x}$Sb$_{x}$ and Bi$_{2}$Se$_3$ \cite{Hasan10,Hasan11}. In contrast, crystal symmetries play a central role in the class of TCIs \cite{Fu11,Slager13}. From a microscopic point of view, necessary conditions for the appearance of metallic surface states on insulating bulk structures comprise general symmetry requirements, strong spin-orbit coupling (SOC), and band inversions in the bulk electronic structure. Concerning TCIs of the face-centered-cubic (fcc, $fm\bar{3}m$) space group, a mirror symmetry causes the appearance of an even number of Dirac cones on (001) and (111) surfaces (which preserve the symmetry; see Fig.~\ref{fig:crystal}a and b). These cones are situated off the TRIMs $\overline{\Gamma}$ and $\overline{\mathrm{X}}$ within the associated high-symmetry line of the surface Brillouin zone \cite{Hsieh12}. Since all five members of the cubic IV-VI class (i.e. GeTe, SnTe, PbS, PbSe and PbTe) belong to the fcc space group, all of them and/or their alloys may be prone to topological phase transitions. Such phase transitions, characterized by a nonzero mirror Chern number \cite{Hsieh12,Teo08}, have been proposed for PbTe and PbSe under pressure \cite{Svane10,Hsieh12}; moreover, they were experimentally observed in ternary alloys Pb$_{1-x}$Sn$_{x}$\ion{A} (\ion{A} = Se, Te) as a function of doping \cite{Dziawa12,Wojek13,Xu12,Tanaka13}. These findings are not simply explained by the strength of SOC: although SOC in Pb ($Z = 82$) is larger than in Sn ($Z = 50$), PbTe is nonetheless a conventional band insulator, as opposed to the TCI SnTe. This clearly calls for a deeper understanding of the microscopic origin of the band inversions.

\begin{figure}
 \centering
 \includegraphics[width=0.9\columnwidth]{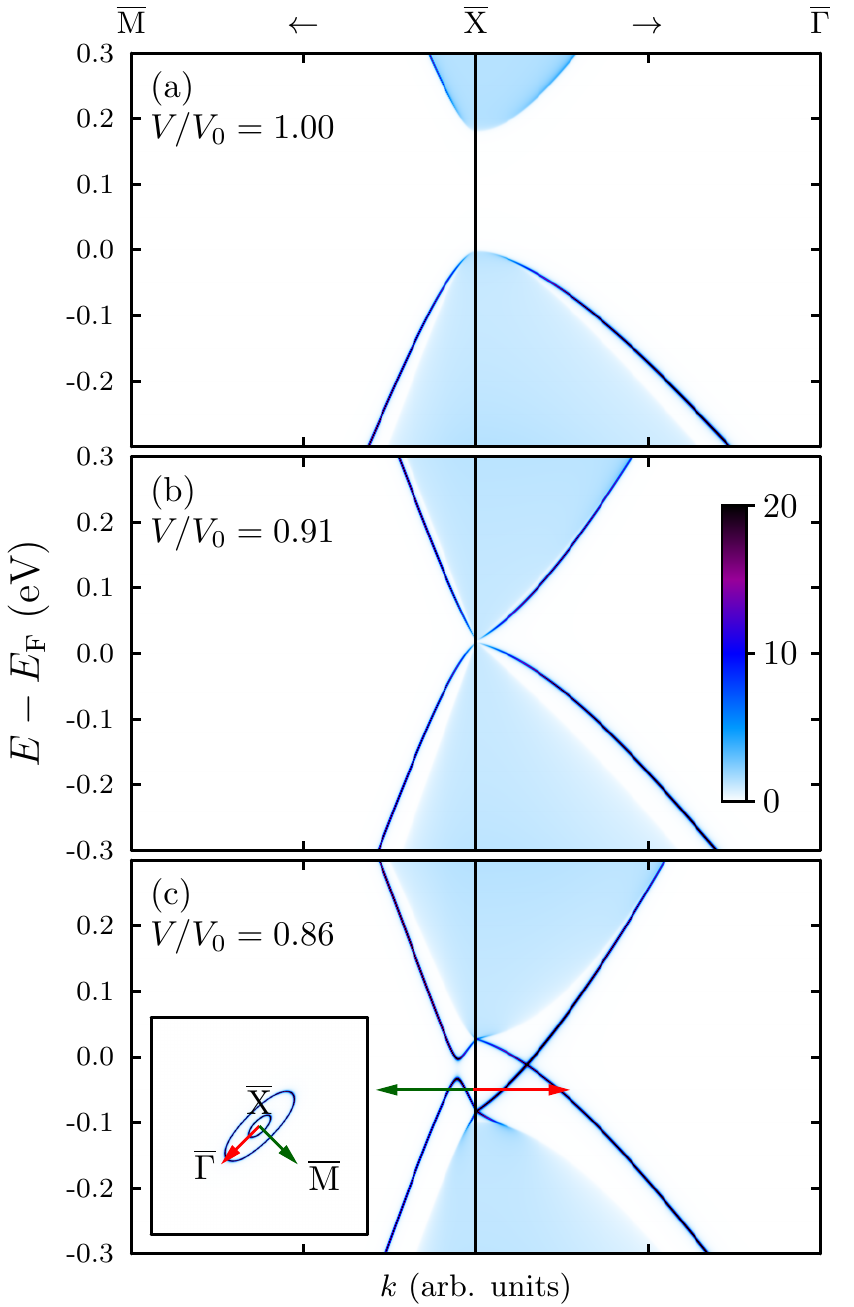}
 \caption{(Color online) Topological phase transition in PbTe upon pressure, calculated by a tight-binding method. For selected volume ratios $\nicefrac{V}{V_{0}}$, the spectral density of the topmost layer of the (001) surface is represented in color scale (b, in $\unit{states / eV}$).  Topological trivial phases with mirror Chern number $0$ appear for the equilibrium volume (a) and close to the critical pressure (b). A topological nontrivial phase (c) shows a mirror Chern number of $-2$. Its Dirac surface state is depicted in a constant-energy cut around $\overline{\mathrm{X}}$ at $E_{\mathrm{F}} - \unit[0.05]{eV}$ (inset in c). A $k$ path shown also in (c) is marked by arrowed lines.  The $k$ axis common to (a--c) shows $\nicefrac{2}{5}$ of the $\overline{\mathrm{M}}$--$\overline{\mathrm{X}}$ and $\overline{\mathrm{X}}$--$\overline{\Gamma}$ lines around $\overline{\mathrm{X}}$ (cfr. Fig.~\ref{fig:crystal}b).}
 \label{fig:bands-tb}
\end{figure}

In this Letter, by means of a semi-empirical analysis, we show that all members of the rock-salt chalcogenide class under consideration can be turned into TCIs under pressure, with the only exception of GeTe. The methods used comprise advanced \textit{first-principles} simulations and \textit{ab initio}-based tight-binding (TB) calculations. Consistent with sophisticated $GW$ computations \cite{Svane10}, we find that the fundamental gap of lead chalcogenides shrinks upon applying external pressure, closes at a critical pressure, and subsequently re-opens with an inverted band character. This behavior is typical of a topological phase transition and exemplified for PbTe in Fig.~\ref{fig:bands-tb}. Here, we used a relativistic tight-binding scheme \cite{Lent86} for the semi-infinite system \cite{Henk93b,Boedicker94}. As long as the fcc structure is preserved, these band inversions cause metallic surface states (Fig.~\ref{fig:bands-tb}c).

\begin{figure*}
\centering
\includegraphics[width=0.7\textwidth]{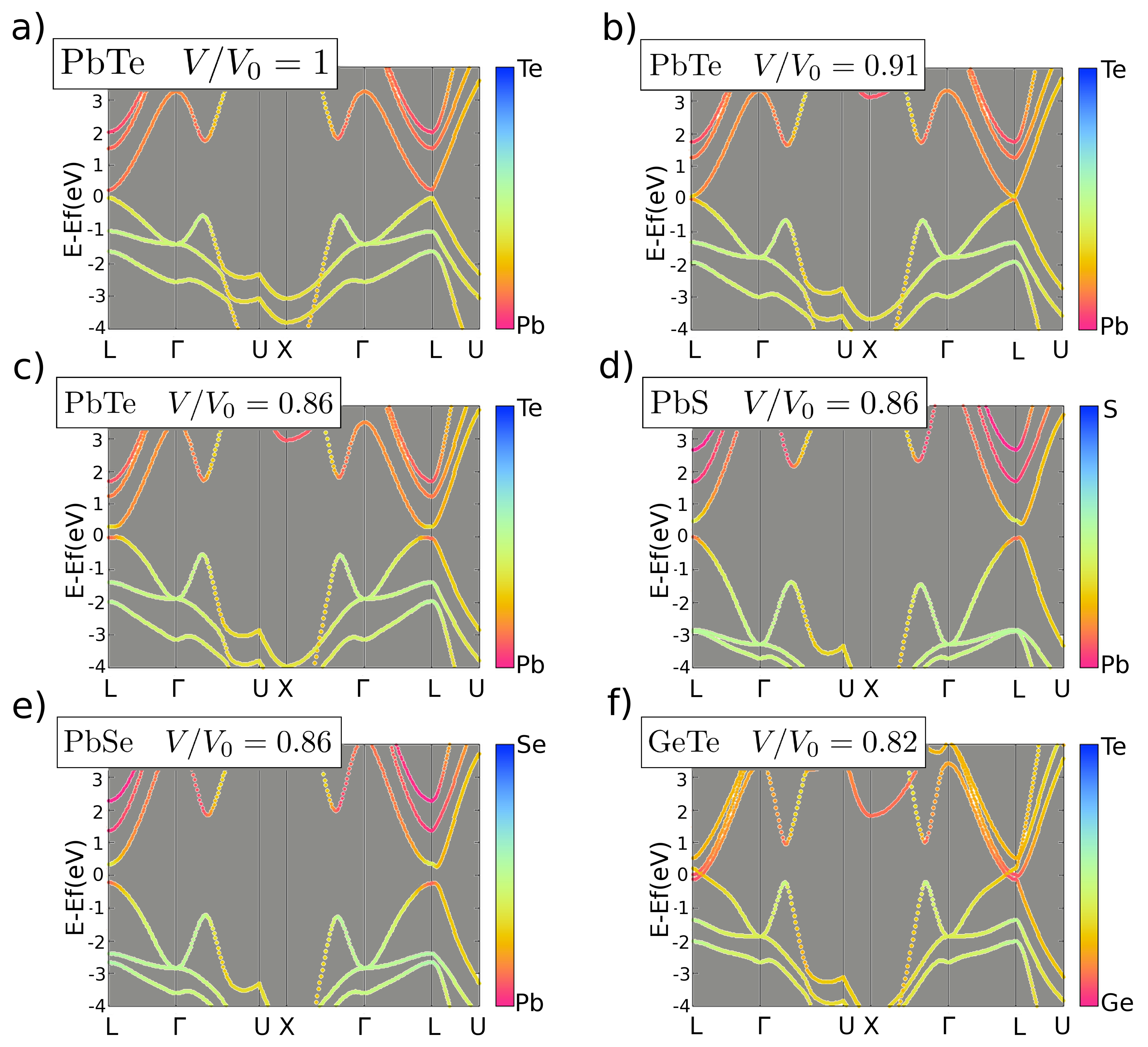}
\caption{(Color online) Electronic structures of selected rock-salt chalcogenides at different pressures (indicated by volume ratios $\nicefrac{V}{V_{0}}$), with band characters highlighted by color scale (anions: blue; cations: red). Simulations are performed within DFT complemented by hybrid functionals.}
\label{fig:bands-vasp}
\end{figure*}

To benchmark the reliability of our predictions, we performed accurate density-functional calculations with hybrid functionals \cite{Heyd04}, as implemented in \textsc{vasp} \cite{Kresse96a,Kresse96b}. These improve significantly with respect to the local-density approximation (LDA) or the (semi-local) generalized-gradient approximation, especially for narrow band-gap semiconductors and lead chalcogenides \cite{Hummer07}. For PbTe, we find the fundamental band gap at L to close at a volume ratio of $\nicefrac{V}{V_{0}} = 0.91$ ($V_{0}$ equilibrium volume; see also Fig.~\ref{fig:bands-tb}b) that is accompanied by band inversion (Fig.~\ref{fig:bands-vasp}). This finding corroborates the topological phase transition deduced from tight-binding calculations (cfr. Fig.~\ref{fig:bands-tb}). At variance with PbTe, hole pockets appear in the electronic structure of GeTe, concomitantly with the pressure-induced closure of the fundamental gap, thus triggering a semi-metallic state instead of a TCI\@. These essential differences require clarification in a microscopic picture.

Assuming the undistorted rock-salt fcc structure, the fundamental band gaps of \ion{CA} compounds are located at eight equivalent L points in the Brillouin zone, at odds with many III-V and II-VI semiconductors in which the direct band gap is at $\Gamma$ \cite{Wei97}. Many peculiar electronic properties of lead chalcogenides Pb\ion{A} were resolved by noting that the Pb-$6s$ band lies below the top of the valence band, thus forming valence bands with the $p$ electrons of the \ion{A} anion. The existence of an occupied cation-$s$ band leads to strong level repulsion at the L point. Due to its symmetry, this explains the narrow band gap $E_{\mathrm{g}}$ and its unusual ordering within the series: $E_{\mathrm{g}}(\mathrm{PbS}) > E_{\mathrm{g}}(\mathrm{PbTe}) > E_{\mathrm{g}}(\mathrm{PbSe})$ \cite{Wei97,Hummer07}. LDA calculations also predicted a negative pressure dependence of the band gap, which would suggest an inverted band structure at L. However, $GW$ calculations have shown that this inversion is a spurious outcome of the local approximation for the exchange-correlation potential, which is known to place the conduction band too low in energy (that is, too small a band gap) \cite{Svane10}. 

The central role of $s$ electrons in determining the electronic properties of \ion{CA} compounds is deduced within the framework of linear combinations of atomic orbitals \cite{Harrison80b}. The only nonzero matrix elements at the L points in the nearest-neighbor approximation are those describing the hybridization between cation (anion) $s$ and anion (cation) $p$ states. Because of SOC, this hybridization close to the Fermi energy involves mainly a combination of atomic-like $p$ states on \ion{C} and \ion{A} sites, with total angular momentum $j = \nicefrac{1}{2}$, namely $\ket{\pm \nicefrac{1}{2}} = \left(\ket{p_{x}, \mp \nicefrac{1}{2}} \pm \cone \ket{p_{y}, \mp \nicefrac{1}{2}} +\ket{p_{z}, \pm \nicefrac{1}{2}}\right) / \sqrt{3}$, and energies $\bar{\epsilon}_{p, \ion{A}(\ion{C})} = \epsilon_{p, \ion{A}(\ion{C})} \pm 2 \lambda_{\ion{A}(\ion{C})}$; here, $\epsilon_{p, \ion{A}(\ion{C})}$ and $\lambda_{\ion{A}(\ion{C})}$ are the orbital energy and the spin-orbit coupling constant of the \ion{C} (\ion{A}) ion, respectively. In the atomic limit that corresponds to the topologically trivial phase, $\Delta_{0} \equiv \bar{\epsilon}_{p, C}-\bar{\epsilon}_{p, A} > 0$ means occupied \ion{A} and unoccupied \ion{C} $p$ shells (Fig.~\ref{fig:crystal}c). The effect of the hybridization with the $s$ states is to push both \ion{C} and \ion{A} $p$ levels to higher energies; this energy shift is roughly proportional to the square of the effective $sp$ hopping interaction $t_{sp}$ and inversely proportional to the energy differences $\Delta_{1} \equiv \bar{\epsilon}_{p, \ion{C}} - \epsilon_{s, \ion{A}}$ and $\Delta_{2} \equiv \bar{\epsilon}_{p, \ion{A}} - \epsilon_{s, \ion{C}}$. The energy gap at L can then be approximated as $E_{g} \approx \Delta_{0} + 10 t_{sp}^{2} \left(\Delta_{1}^{-1} - \Delta_{2}^{-1}\right) / 3$. Hence, for band inversion to occur, two conditions have to  be satisfied: 1. the energy separation between atomic-like $p$ states (including SOC) must be sufficiently small, and 2. the $sp$ hybridization should be strongly ``asymmetric''. This asymmetry is realized if $\Delta_{1} \gg \Delta_{2}$, i.\,e.\ if the cation (anion) $s$ level is energetically close to (far from) the anion (cation) $p$ states.

The band gap $E_{\mathrm{g}}$ is directly related to the $\vec{k} \cdot \vec{p}$ Hamiltonian at L, $\widehat{H} = m \sigma_{z}$ \cite{Mitchell66,Hsieh12}; $\sigma_{z} = \pm 1$ corresponds to the $p$ character on cation \ion{C} and anion \ion{A} sites, respectively. Furthermore, a negative $m \equiv E_{\mathrm{g}} / 2$ implies that conduction and valence bands at L would respectively derive from \ion{A} and \ion{C} ions. A reversal of $m$ in the presence of the mirror symmetries in the fcc structure would imply a topological phase transition with an associated change of the mirror Chern number \cite{Hsieh12}. Since $\Delta_{0} = \epsilon_{p, \ion{C}} - \epsilon_{p, \ion{A}} - 2 (\lambda_{\ion{C}} + \lambda_{\ion{A}})$, a large SOC is required to fulfill the first condition. The second condition, on the other hand, could in principle be controlled by alloying ternary solid solutions \ion{C}$_{x}$\ion{C'}$_{1-x}$\ion{A} (along the path pursued in \cite{Dziawa12,Wojek13,Xu12,Tanaka13}) or $\ion{CA}_{x}\ion{A'}_{1-x}$, since all energy differences are expected to change: within the virtual-crystal approximation (VCA) one has $\Delta_{i} = x \Delta_{i}^{\ion{CA}} + (1-x) \Delta_{i}^{\ion{C'A} (\ion{CA'})}$. Because the band inversion is proportional to the strength of the $sp$ hybridization, a straightforward way to induce a topological phase transition is to apply external pressure, as this directly affects the $sp$ hopping $t_{sp} \propto d^{-2}$ \cite{Harrison80b,Hsieh12,Svane10}.

Guided by these considerations, we  performed an empirical screening within the family of rock-salt
chalcogenides, using both Harrison's \cite{Harrison80b} and Lent's \cite{Lent86} parameterizations
(Table~\ref{tab:differences}). Even if figures are different, trends are consistent within both parameterizations.
We can therefore loosely identify three subclasses: (i) The first subclass --- that comprises PbTe and SnTe --- is characterized by a relatively small $\Delta_{0}$ (mainly due to the large $\lambda_{\mathrm{Te}}$) and similar differences $\Delta_{1}^{-1} - \Delta_{2}^{-1}$. In this respect, the main reason why SnTe is a TCI but PbTe is not, could be the smaller equilibrium volume of the former with respect to the latter. In turn, a relatively small pressure could tune the topological transition in PbTe \cite{Hsieh12} and the doping-dependent topological transition in Pb$_{1-x}$Sn$_{x}$Te could be ascribed to a (chemical) pressure effect. (ii) The second subclass comprises PbS and PbSe and shows $\Delta_{0}$'s approximatively twice as large as those of the first subclass, due to the smaller $\lambda_{\ion{A}}$. However, $\Delta_{2}$ is much smaller than $\Delta_{1}$; it is thus very likely that a reduction of the lattice constant and the associated increase of $t_{sp}$ could result in band inversion (due to the strongly asymmetric $sp$ hybridization) and, hence, in a topological phase transition.
(iii) Eventually GeTe, belonging to the third class, even though displaying a small $\Delta_{0}$, does not seem to fulfill condition 2. Thus, a reopening of the gap after a pressure-induced closure is unlikely and a transition to a metallic state (rather than toward a TCI) is expected upon applying pressure (Fig.~\ref{fig:bands-vasp}).

\begin{table}
\caption{Estimated relevant quantities from Harrison's \cite{Harrison80b} and Lent's (\cite{Lent86}, in brackets) parameterization. Spin-orbit coupling constants are taken from atomic values, as estimated in \cite{Montalti06}. Lent's parameterization has been obtained through a fitting procedure including also $d$ states. Energy differences $\Delta$ (in $\unit{eV}$); lattice constants $a_{0}$ (in $\unit{\AA}$) taken from \cite{Abrikosov68}.}
\centering
\begin{tabular}{p{1.3cm}|p{1.3cm}|p{1.3cm}|p{1.3cm}|c}
\hline\hline
&&&&\\[-0.3cm]
	&\centering{$\Delta_{0} $}   & \centering{$\Delta_{1} $}   &\centering{$\Delta_{2} $}	&$a_{0}$ \\
\hline
&&&&\\[-0.3cm]
\centering{PbTe } 	&\centering{0.71}	&\centering{10.08}	&\centering 4.32& 6.462	\\
 	&\centering{(1.72)}	&\centering{(12.70)}	& \centering{(7.16)}	&\\[0.1cm]
\centering{SnTe }	&\centering{1.35}		&\centering{10.71}	&\centering4.75	&6.327	\\
 	&\centering{(0.95)}	&\centering{(13.13)}	&\centering{(6.69)}		& \\
\hline
&&&&\\[-0.3cm]
\centering{PbS	}&\centering{3.15}	&\centering{13.77}		&\centering1.89		&5.936	\\
 	&\centering{(3.92)}	& \centering{(16.25)}	&\centering(5.59)	&\\[0.1cm]
\centering{PbSe}	&\centering{2.09}		&\centering13.29	&\centering 2.95		&6.124 	\\
 	&\centering{(3.29)}		&\centering{(15.75)}	&\centering(5.18)	& \\
 \hline
 &&&&\\[-0.3cm]
\centering{GeTe}	&\centering{1.19}		&\centering{10.55}		&\centering6.63		&6.009	\\
 	&\centering{(0.28)}	&\centering{(11.92)}	&\centering (8.51)		& \\
\hline\hline
\end{tabular}
\label{tab:differences}
\end{table}

To verify the band inversions at the L points, we have performed \textit{ab initio} hybrid-functional calculations
\cite{Heyd04}\footnote{Kohn-Sham equations were solved using the projector augmented-wave method. The energy
cutoff for the plane-wave expansion was $\unit[600]{eV}$; an $8\times 8 \times 8$ Monkhorst-Pack $k$-point grid
 was used. Calculations with the hybrid HSE functional \cite{Heyd04} are computationally very demanding and, thus,
 were used for bulk states only.}.  The band inversion is highlighted for PbTe, PbS, PbSe, and GeTe at selected
 volumes (Fig.~\ref{fig:bands-vasp}). For PbTe, the closure of the band gap shows up at a critical volume
 of $0.92 V_{0}$; hence, the band inversion has already occurred at a volume ratio $\nicefrac{V}{V_{0}} = 0.91$
 (pressure of about $\unit[4]{GPa}$; our TB calculations give the same critical volume, Fig.~\ref{fig:bands-tb}).
 It shows a predominant Te anion character in the conduction band at the L point; the negative gap increases upon
 further decreasing the volume ($\nicefrac{V}{V_{0}} = 0.86$ in Fig.~\ref{fig:bands-vasp}). Indeed,
 the atomic and orbital resolved character of the valence band maximum (conduction band minimum) at L shows a
 predominant Pb-$s$ and Te-$p$ (Pb-$p$ and Te-$s$) contribution at the equilibrium volume, whereas an opposite
 character appears as the volume is decreased below the critical value\footnote{When projecting wavefunctions at
 L within muffin-tin spheres with radius = 1.73 \AA\, and 1.54 \AA\, for Pb and Te respectively, the weights on Pb-$s$,
 Pb-$p$, Te-$s$, and Te-$p$ are $0.24$, $0.0$, $0.0$, and $0.36$ for the valence band maximum (VBM), and $0.0$, $0.36$,
 $0.06$, $0.0$ for the conduction band minimum (CBM) at the equilibrium volume for PbTe, whereas we find $0$, $0.36$,
 $0.06$, $0$ for the VBM and $0.29$, $0.0$ $0.0$, $0.11$ for the CBM  at $\nicefrac{V}{V_{0}} = 0.86$.}. The same trend
 with pressure holds for both PbS and PbSe, with critical volumes of about $0.94 V_{0}$ and $0.96 V_{0}$ (tight-binding:
 $0.91 V_{0}$ and $0.96 V_{0}$), respectively, corresponding to a pressure of about $\unit[4.2]{GPa}$ ($\unit[6.3]{GPa}$) for  PbS and  $\unit[2.4]{GPa}$ for PbSe. For GeTe, on the other hand, the
 gap closes at a critical volume of $0.9 V_{0}$, corresponding to a rather large pressure of $\unit[5]{GPa}$. This finding is consistent with the smaller difference between $\Delta_{1}$ and $\Delta_{2}$ as compared to those of PbTe. Furthermore, since the $j = \nicefrac{1}{2}$ and $j = \nicefrac{3}{2}$ manifolds in the conduction bands are close in energy due to the small SOC in Ge, a further increase of pressure would push them both below the anion $p$ states, thus turning GeTe to a semimetal (Table~\ref{tab:differences} and Fig.~\ref{fig:bands-vasp}).

To provide further support for the pressure-induced topological phase transitions, we performed relativistic tight-binding calculations for bulk PbTe, PbSe, PbS, and GeTe. The computation of the mirror Chern number is done in the spirit of the spin Chern number \cite{Prodan09}. Since the Bloch states are eigenstates of both the Hamiltonian and the mirror operator \cite{Inui90}, we separate the Bloch states into two categories with mirror eigenvalues $\pm \cone$ and calculate the  Berry curvature for both. The integral of the Berry curvature over the intersection of the mirror plane with the Brillouin zone yields Chern numbers $n_{\pm \cone}$, from which the mirror Chern number $c_{\mathrm{m}} \equiv (n_{+\cone} - n_{-\cone}) / 2$ is obtained  \cite{Teo08}. While GeTe is always topologically trivial, we find a mirror Chern number of $-2$ for band-inverted PbTe, PbSe, and PbS. In summary, the numerical calculations corroborate the microscopic picture derived above.

In conclusion, we have shown that a strongly asymmetric hybridization between cation (anion) $s$ and anion
(cation) $p$ states, together with a sizeable strength of spin-orbit coupling, is a necessary condition for band
inversions to occur at the L points and the related topological crystalline state to arise in rock-salt chalcogenides.
By performing a thorough analysis of pressure effects in the entire family of fcc chalcogenides, we verify the
topological nature of the transition, as shown by nonzero mirror Chern numbers and gapless edge states at (001)
surfaces.  Furthermore, we suggest that lead chalcogenides with topologically trivial properties could be
turned into a topologically nontrivial state upon a combination of applied pressure and
alloying \emph{through anion substitution}. For instance, in the PbSe$_{x}$Te$_{1-x}$ alloy
one expects an effect of chemical pressure
($a_{0}(\mathrm{PbSe}) \ll a_{0}(\mathrm{PbTe})$) and of the band shifts. The latter ``asymmetrize''
the hopping interactions, so that for a doping of $x = 0.2$ one would get $\Delta_{0} = \unit[1]{eV}$,
$\Delta_{1} = \unit[10.72]{eV}$, $\Delta_{2} = \unit[4.04]{eV}$, and $a_{0} = \unit[6.39]{\AA}$ in the VCA\@.
Also on the basis of the smaller critical pressure predicted for PbSe, a 
topological transition is therefore likely to occur at reasonable values of applied pressure. 
In this context, we remark that alloying two
topologically trivial insulators is a yet unexplored path to engineer a TCI.
This scenario
may be relevant for the experimental search of conducting edge states, as the carrier concentration, largely determined
by the presence of cation/anion vacancies, can be easily controlled  in lead
chalcogenides during the crystal growth\cite{Khokhlov03}. On the other hand,
a high concentration of cation vacancies
is frequently found in binary SnTe, resulting in an undesirable $p$-type degenerate conducting behavior,
a drawback that could be overcome by means of Pb$A$ anion alloying.

\acknowledgments
This work is supported by the Priority Program 1666 `Topological Insulators' of the DFG\@.
We acknowledge PRACE for awarding us access to resource MareNostrum based
in Spain at Barcelona Supercomputing Center (BSC-CNS).

\bibliographystyle{apsrev}

\end{document}